\documentclass[runningheads,a4paper]{llncs}

\usepackage{amssymb}
\usepackage{graphicx}
\usepackage{url}
\usepackage{latexsym}
\usepackage{graphicx}
\usepackage{mathrsfs}
\usepackage{amsmath}
\usepackage{multirow}
\usepackage{epstopdf}
\usepackage{caption}
\usepackage {algorithm2e}
\usepackage{lipsum}
\usepackage{cite}

\urldef{\mailsa}\path|{tiennm, anhphanviet,nguyenml}@jaist.ac.jp|
\urldef{\mailsb}\path|anna.kramer, leonie.kunz, christine.reiss, nicole.sator,|
\urldef{\mailsc}\path|erika.siebert-cole, peter.strasser, lncs}@springer.com|    
\newcommand{\keywords}[1]{\par\addvspace\baselineskip
\noindent\keywordname\enspace\ignorespaces#1}

\begin{document}

\mainmatter  

\title{Learning to Rank Questions for Community Question Answering with Ranking SVM}

\titlerunning{QCRI at the ECML/PKDD 2016 Challenge on Question Re-Ranking}

%
%
\author{%
  Minh-Tien Nguyen$^{1, 2}$, 
  Viet-Anh Phan$^1$,
  Truong-Son Nguyen$^1$ and\\
  Minh-Le Nguyen$^{1}$
}
\authorrunning{Nguyen, Phan, Nguyen, and Nguyen}

\institute{%
	 $^1$ School of Information Science,\\
	Japan Advanced Institute of Science and Technology (JAIST),\\
  1-1, Asahidai, Nomi, Ishikawa, 923-1292, Japan.\\
  $^2$ Hung Yen University of Technology and Education (UTEHY), Vietnam.
\mailsa\\
}

%
%

\maketitle

\begin{abstract}
This paper presents our method to retrieve relevant queries given a new question in the context of Discovery Challenge: Learning to Re-Ranking Questions for Community Question Answering competition. In order to do that, a set of learning to rank methods was investigated to select an appropriate method. The selected method was optimized on training data by using a search strategy. After optimizing, the method was applied to development and test set. Results from the competition indicate that the performance of our method outperforms almost participants and show that Ranking SVM is efficient for retrieving relevant queries in community question answering.

\end{abstract}

\section{Introduction}\label{sec:introduction}
\begin{figure}
	\includegraphics[scale=1]{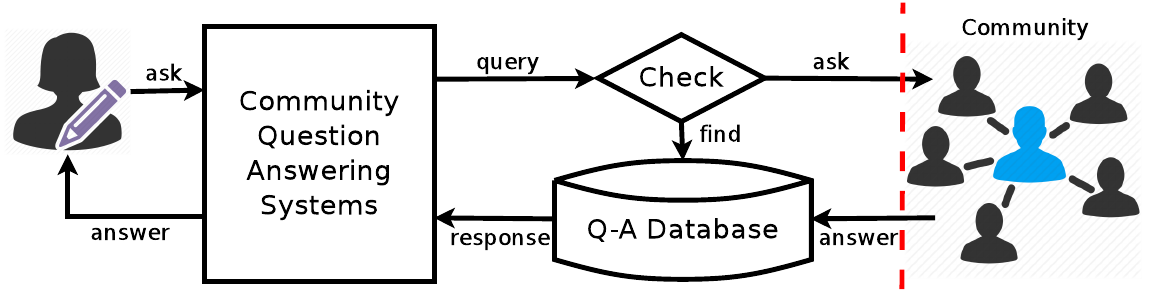}
    \caption{The illustration of an QA-community system}\label{fig:example}\vspace{-0.6cm}
\end{figure}

\vspace{-0.2cm}
The rapid growth of textual data has inspired the development of retrieval systems in which users put a query or question and expect to obtain relevant data corresponding to the query or question. For example, in a Community Question Answering (cQA) system, e.g. Stackoverflow\footnote{http://stackoverflow.com} or Yahoo! Answer\footnote{https://answers.yahoo.com}, whenever a user poses a new question, the user hopefully obtains related answers from the system. If the new question is similar to a previously posted question (even semantically equivalent), the c-QA in Fig. \ref{fig:example} should return the question-forum thread to response the new question. An automatic retrieval system can search for the previously-posted question and provide relevant question-forum thread for the new question. This system benefits users as well as experts in which users receive an answer immediately and experts do not need to response.

In the context of Discovery Challenge: Learning to Re-Ranking Questions for Community Question Answering\footnote{http://alt.qcri.org/ecml2016/index.html}, given a new user question and a set of previously-posted-forum questions, together with their corresponding answer threads, a machine learning system has to rank the forum questions according to their relevance to the new question. Formally, given a new question $u$, a set of previously-posted-question-forum threads $(q_i, A_j)$, in which $q_i$ is a previous question ($i \in \{1, Q\}$) and $A_j$ contains a set of posted answers from other users corresponding to each $q_i$, the mission of the competition is to find \emph{\textbf{the best match}} of $q_i$ giving $u$.

Learning to rank (L2R) for information retrieval has been received a lot of attention from the research community. Ni et al. proposed a new ranking approach for information retrieval, where the diversity among queries was taken into consideration \cite{NHX-WAIM-08}. The authors treated the probability distribution of retrieved documents by relaxing assumption, e.g. independently and identically distributed to fit the real situations of information retrieval. Huang et al. proposed a Bayesian learning approach to promoting diversity for information retrieval in biomedicine and a re-ranking model to improve retrieval performance in biomedical domain \cite{HH-SIGIR-09}. Faria et al. addressed the problem of retrieving images relying on content-based approach \cite{FVAVTAGM-MIR-10} and proposed a L2R method based on Support Vector Machine, genetic programming, and association rules. Azarbonyad et al. investigated the cross-language information retrieval in which input was a query of a source language, a retrieval system had to return relevant information in a target language \cite{ASF-ECAI-12}. In document summarization, Svore et al. proposed a method which used additional features extracted from Wikipedia to enrich the information of a sentence and applied RankNet \cite{BSRLDHH-ICML-05} to summarize single-document \cite{SVB-EMNLP-CoNLL-07}. Wei and Gao proposed a L2R model which exploited the support from tweets to summarize single-document \cite{WG-COLLING-14}. Nguyen et al. used Ranking SVM to summarize single-document on their dataset \cite{NTTN-CIKM-16}.

The objective of this paper is to select and optimize an appropriate L2R method on the data of the competition. In order to do that, a set of L2R methods is first analyzed to select an appropriate one. This method is then applied to development and test set. This paper also reports the results of the competition with discussions. This paper finally draws important conclusions.\vspace{-0.2cm}

\section{Ranking Questions for Community Question Answering}\label{sec:method}\vspace{-0.3cm}
This section shows our proposal to address the task of the competition in three steps: basic idea, data preparation and learning to rank with Ranking SVM.\vspace{-0.4cm}

\subsection{Basic Idea}
\begin{figure}
	\includegraphics[scale=1]{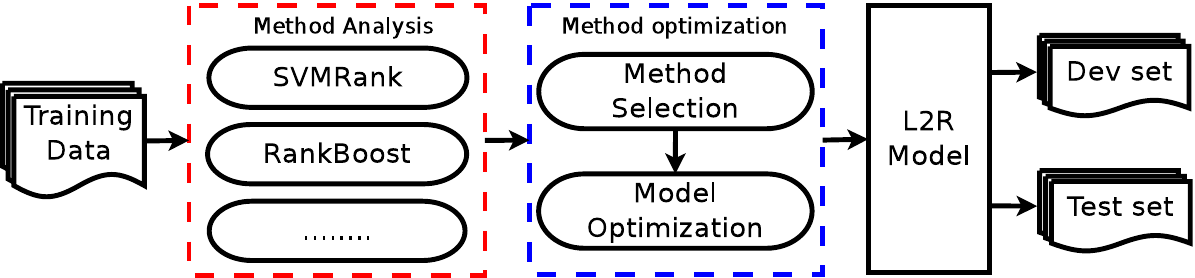}
    \caption{The overview of our method}\label{fig:model}\vspace{-1cm}
\end{figure}

\vspace{-0.2cm}
In the context of the competition, our idea is to find a L2R method and optimize the performance of the method. In order to do that, a set of L2R methods is first analyzed. The analysis helps to figure out an appropriate method. Next, the selected method is optimized by adjusting hyper-parameters so that they maximize the performance of the method  on the training set. Subsequently, a model is trained with the best hyperparameters on the training data and applied on development and test data. Our method is shown in Fig. \ref{fig:model}.\vspace{-0.1cm}

\subsection{Data Preparation}\vspace{-0.2cm}
The data of the competition contained three sets: training, development and test, in which the development, and test set had no labels. The training set was used to train a learning to rank model and the development and test set were used to test the performance of the model. Each question was paired with 10 answers. The statistics of data is shown in Table \ref{tab:data}.\vspace{-0.5cm}
\begin{table}[ht]
\centering
\caption{Data statistics}\label{tab:data}
\begin{tabular}{l c c}
\multicolumn{1}{c}{\textbf{Set name}} & \textbf{Questions} & \textbf{Question-answer pairs} \\ \hline
Training                              & 267               & 2670                  \\
Development                           & 50                & 500                   \\
Test                                  & 70                & 700            \\ \hline      
\end{tabular}\vspace{-0.3cm}
\end{table}

In order to analyze L2R methods, the original training data was split into two sets: sub-training and sub-test. The sub-test was generated by picking up last 50 questions corresponding to 500 query-answer pairs in the original training data and therefore, the sub-training contained 217 questions corresponding to 2170 question-answer pairs.

\subsection{Experimental Setup}
The sub-training was used to train and sub-test was used to test L2R methods. Due to the aim of finding an appropriate method, default parameters of each method were used without tuning parameters or feature selection. The development and test set were used to compare the performance of our method with other participant teams in the competition.

The mean of all the average precision values (MAP)\footnote{https://en.wikipedia.org/wiki/Information\_retrieval\#Mean\_average\_precision} was used to compare L2R methods. MAP for a set of queries was defined by the mean of the average precision for each query and shown in Eq. \eqref{eq:map}.
\begin{equation}\label{eq:map}
	MAP = \frac{\sum_{q=1}^Q AveP(q)}{Q}
\end{equation}
where: \emph{Q} is the number of queries. $AveP()$ returns the average precision of a ranked sequence items corresponding to a query. Note that although the competition provided a set of information retrieval metrics, e.g. F-1 or Accuracy, MAP was formally used as a major metric.\vspace{-0.2cm}

\subsection{Learn to Rank Approaches}
\subsubsection{Feature Extraction:}
Due to the aim of focusing on machine learning aspects, the data was presented in the form of vectors including 64 state-of-the-art features. The features were divided into three sub-groups.
\begin{itemize}
	\item Similarity aspect: included 47 features which evaluate similarities between a new and its related forum questions.
    \item Related aspect: contained four features representing whether the forum question is strongly related to a new question. If the forum thread contains good answers to the new question, an Answer Selection Classifier was applied to estimate the quality of comments from the forum question thread with respect to the new question.
    \item Classification aspect: contained 13 features for evaluating the discrepancy between comments from the forum question thread with respect to both the new and the forum question. To generate such scores, an Answer Selection Classifier was applied.
\end{itemize}

The features were provided in the form of \emph{LibSVM}\footnote{https://www.csie.ntu.edu.tw/~cjlin/libsvm/} format. In order to apply for learning to rank methods, feature vectors were converted into the form of SVMRank\footnote{https://www.cs.cornell.edu/people/tj/svm\_light/svm\_rank.html}. Note that the format of SVMRank was also used for RankLib\footnote{https://people.cs.umass.edu/$\sim$vdang/ranklib.html}.

\subsubsection{Method Analysis:}
A set of L2R methods was investigated to select an appropriate method. Given $n$ training queries $\{q_i\}_{i=1}^n$, their associated document pairs $(x_u^{(i)}, x_v^{(i)})$ and the corresponding ground truth label $y_(u,v)^{(i)}$, the methods are listed as the following:
\begin{itemize}
	\item \textbf{Ranking SVM}: applies the characteristics of SVM to perform pairwise classification \cite{J-KDD-06}, which optimizes the objective function shown in Eq. \eqref{eq:objective-function}:
\begin{align}\label{eq:objective-function}
		& \min \frac{1}{2} \left \| w \right \|^2 + \lambda \sum_{i=1}^n \sum_{u, v:y_{u,v}^{(i)}} \xi_{u, v}^{(i)}\\
		\text{s.t. } & w^T (x_u^{i} - x_v^{(i)}) \geqslant 1 - \xi_{u, v}^{(i)} \text{, if } y_{u, v}^{(i)} = 1\\
    	& \xi_{u, v}^{(i)} \geqslant 0 \text{,   } i = 1,..., n
\end{align}
where $f(x) = w^Tx$ is a linear scoring function, $(x_u, x_v)$ is a pairwise and $\xi_{u, v}^{(i)}$ is the loss.
    To test this method, SVMRank\footnote{https://www.cs.cornell.edu/people/tj/svm\_light/svm\_rank.html} was used with \emph{linear} kernel and \emph{C=3}.
    
    \item \textbf{RankBoost}\cite{FLSS-ICML-98,FLSS-JMLR-03}: was adopted from AdaBoost \cite{YS-ECCL-95} for the classification over document pairs. The only difference between RankBoost and AdaBoost is that the distribution in RankBoost was defined on document pairs whereas, in AdaBoost, the distribution was defined on individual documents \cite{L-Springer-11}. In training, RankBoost maximizes an exponential loss defined in Eq. \eqref{eq:rankboost}.
\begin{equation}\label{eq:rankboost}
	L(f; x_u, x_v, y_{u, v}) = exp (-y_{u, v}(f(x_u) - f(x_v)))
\end{equation}
   RankBoost was used with \textit{iteration} = 300, \textit{metric} is ERR\@10.
    
    \item \textbf{RankNet}\cite{BSRLDHH-ICML-05}: defined the loss function on a pair of documents, but the hypothesis was defined by as the cross entropy loss shown in Eq. \eqref{eq:ranknet}.
\begin{equation}\label{eq:ranknet}
	L(f; x_u, x_v, y_{u, v}) = -\bar{P}_{u, v} log P_{u, v} (f) - (1-\bar{P}_{u, v}) log (1-P_{u, v} (f))
\end{equation}
where: $P_{u, v}$ is the modeled probability; a target probability $\bar{P}_{u, v}$ was constructed based on ground truth labels, i.e. $\bar{P}_{u, v} = 1$ if $y_{u, v} = 1$. RankNet was applied with \emph{epoch} = 100, \textit{the number of layers} = 1, \textit{the number of hidden nodes} per layer = 10 and l\textit{earning rate} = 0.00005.

    \item \textbf{AdaRank}\cite{XL-SIGIR-07}: directly optimized evaluation measures using non-smooth optimization techniques. In AdaRank, the evaluation measures were used to update query distribution and to compute the combination coefficient of weak rankers. AdaRank was applied with \textit{the number of training rounds} = 500, \textit{tolerance between two consecutive rounds} of learning = 0.002, and \textit{the maximum number of times} can a feature be consecutively selected without changing performance = 5.
    
    \item \textbf{Radom Forest}\cite{B-ML-01}: combined tree predictors in which each tree depends on the values of a random vector sampled independently and with the same distribution for all trees in the forest. Random Forest was applied with \textit{the number of bags} = 300, \textit{sub-sampling rate} = 1.0, \textit{feature sampling rate} = 0.3, \textit{ranker to bag with MART}, \textit{the number of trees} in each bag = 100, \textit{learning rate} = 0.1, and \textit{the min leaf support} = 1.
\end{itemize}

Note that in training, the pair-wise order is the label of each instance in the training dataset. RankBoost, RankNet, AdaRank, and Random Forest were used in RankLib\footnote{https://people.cs.umass.edu/$\sim$vdang/ranklib.html}.\vspace{-0.3cm}
\begin{table}[ht]
\centering
\caption{Learning to rank results on the sub-test set}
\label{tab:l2r-result}
\begin{tabular}{c p{1cm} p{1cm} p{1cm} p{1cm}}
\hline
\multirow{2}{*}{\textbf{L2R Method}} & \multicolumn{4}{c}{\textbf{Sub-Test Set}} \\ \cline{2-5} 
                                 & \textbf{MAP}       & MRR      & P@1      & P@5      \\ \hline
SVMRank                          & \textbf{0.7456}    & 0.8388   & 0.8200   & 0.5640   \\
RankBoost                        & 0.7201    & 0.7858   & 0.7200   & 0.5520   \\
RankNet                          & 0.6317    & 0.6741   & 0.5600   & 0.5000   \\
AdaRank                          & 0.6503    & 0.7306   & 0.6400   & 0.5280   \\
Random Forest                    & 0.7042    & 0.7761   & 0.7400   & 0.5280   \\ \hline
\end{tabular}\vspace{-0.3cm}
\end{table}

The results of these methods on the sub-test set in Table \ref{tab:l2r-result} indicate that SVMRank is the best method, i.e. MAP\footnote{Although MAP was officially used for evaluation, other metrics were also reported.} = 0.7456, RankBoost and Random Forest are competitive methods while RankNet and AdaRank obtain quite poor results. From the observation, SVMRank was selected to apply in the development and real test set.\vspace{-0.4cm}

\subsubsection{Model Optimization:}
To optimize SVMRank, two steps were considered: selecting kernel and tuning hyperparameters. In the first step, linear, radial basis function (RBF) and sigmoid tanh were used with $C=3$. Other default hyper-parameters were also used. Results in Table \ref{tab:kernel} indicate that SVMRank with linear kernel obtains the best result. As the result, \textit{linear} kernel was selected for training SVMRank.
\begin{table}[ht]
\centering
\caption{The performance of SVMRank with various kernel}
\label{tab:kernel}
\begin{tabular}{c c c c c}
\hline
\multirow{2}{*}{\textbf{Kernel}} & \multicolumn{4}{c}{\textbf{Sub-Test Set}} \\ \cline{2-5} 
                                 & \textbf{Map}       & MRR      & P@1      & P@5      \\
Linear                           & 0.7456    & 0.8388   & 0.8200   & 0.5640   \\
RBF                              & 0.6620       & 0.8388      & 0.8200      & 0.5640      \\
Sigmoid                          & 0.5327       & 0.5493      & 0.3880      & 0.4480     \\ \hline
\end{tabular}\vspace{-0.3cm}
\end{table}

In tuning hyperparameters, since SVMRank with \textit{linear} kernel has only one parameter $C$, which is the trade-off of training error, our mission is to find out an appropriate value which optimizes the performance of the model. The selection was conducted in two steps: range finding and value selecting. In the first step, $C$ was empirically tested with various values \{3, 30, 300, 300, 3000, 30.000\}. This step figures out the range of $C$ which the model performs well.
\begin{table}[ht]
\centering
\caption{The performance of SVMRank with various range of $C$}
\label{tab:c}
\begin{tabular}{c c c c c}
\hline
\multirow{2}{*}{\textbf{Trade-off parameter}} & \multicolumn{4}{c}{\textbf{Sub-Test Set}} \\ \cline{2-5} 
   & \textbf{Map}       & MRR      & P@1      & P@5      \\
3          & 0.7456    & 0.8388   & 0.8200   & 0.5640   \\
30         & 0.7417       & 0.8291      & 0.8000      & 0.5600      \\
300       & 0.7336       & 0.8075      & 0.7600      & 0.5680      \\
3000      & 0.7384       & 0.8075      & 0.7600      & 0.5720     \\
30.000     & 0.7314       & 0.7941      & 0.7400      & 0.5680     \\ \hline
\end{tabular}\vspace{-0.2cm}
\end{table}

Results in Table \ref{tab:c} show that with $C \in [3, 30]$, the model is the best. After finding a suitable range, an appropriate $C$ value must be selected in this range. To find the value, $C$ was tuned in [3, 30], in which $C=3$ is the initial value. Two additional $C$ values, i.e. 35 and 40 were also tested to show the trend of MAP. The selection was conducted by using a searching mechanism in Algorithm \ref{alg:searching}.

\begin{figure}[ht]
	\centering
	\includegraphics[scale=0.6]{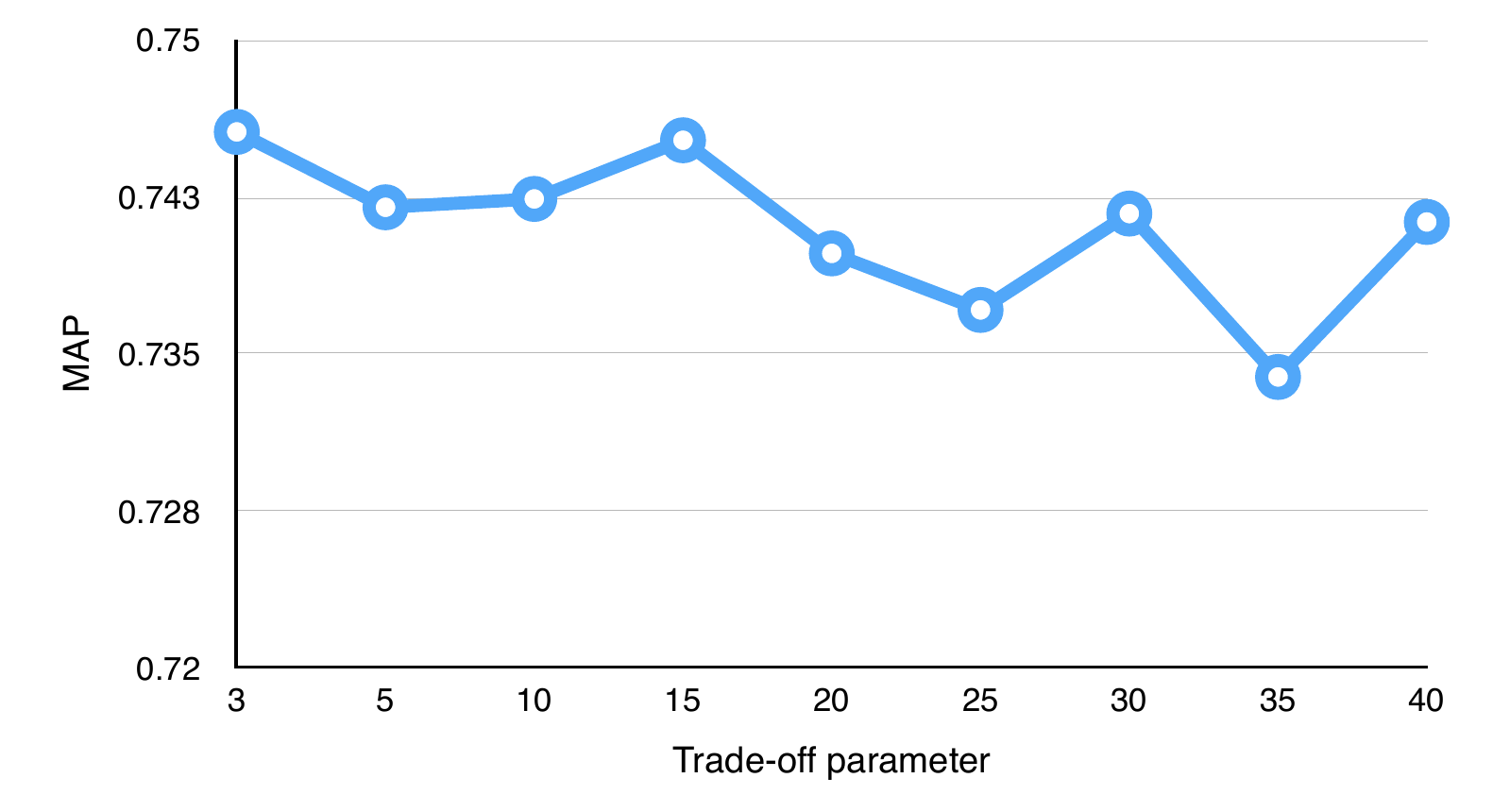}\vspace{-0.1cm}
    \caption{The performance of SVMRank with various $C \in [3, 40]$}\label{fig:c}\vspace{-0.3cm}
\end{figure}

\begin{algorithm}[ht]
 \KwData{\textit{linear} kernel and $C$ in [5, 40]}
 \KwResult{an appropriate $C$ value}
 \textit{MAP-tmp} = the MAP of SVMRank with $C=3$ (initial step)\;
 \For{c=5 to 40}{
  \textit{MAP} = \textit{MAP} of SVMRank with the current $c$\;
  \If{MAP-tmp$<$MAP}{
   $C$=$c$\;
   }
   $c$ = $c$ + $step$\;
 }
 \caption{Searching algorithm}\label{alg:searching}
\end{algorithm}

where: $step=5$. Algorithm \ref{alg:searching} first loops each $C$ value in [5, 40], in each value, the algorithm calculates the MAP of SVMRank, if this value is larger than the initial MAP value, then the current $C$ value is stored. In Fig. \ref{fig:c}, MAP score decreases with $C$ in [3, 15] and reach the top at $C=15$. After that, MAP score tends to decrease again.

Results from Table \ref{tab:c} and Fig. \ref{fig:c} conclude that our method is the best at $C=15$ with \textit{linear} kernel.

\section{Competition Results and Discussion}\label{sec:result-discussion}\vspace{-0.1cm}
The results of the competition in Table \ref{tab:result} indicate that \emph{unocanda} is the best on both development and test set. As the result, \emph{unocanda} achieved the first prize in the competition. Our method named \emph{sakuraSAT} obtained the third position on the both datasets. \emph{MI} team got the second rank on test set; however, they did not meet the requirements of the competition. As the result, our team obtained the second prize of the competition.  Note that our method significantly outperforms the baseline and about over sixty teams who registered the competition.
\begin{table}[ht]
\centering
\caption{Competition results of top six teams on development and test set}
\label{tab:result}
\begin{tabular}{ccccc|cccc} \hline
\multirow{2}{*}{\textbf{Rank}} & \multirow{2}{*}{\textbf{Name}} & \multicolumn{3}{c|}{\textbf{Dev Set}} & \multirow{2}{*}{\textbf{Name}} & \multicolumn{3}{c}{\textbf{Test Set}} \\ \cline{3-5} \cline{7-9} 
 &   & \textbf{Map}        & AvgREC     & MRR        &  & \textbf{MAP}        & AvgREC     & MRR         \\ \hline
1  & unocanda                            & 0.7536     & 0.8996     & 79.888     & unocanda                            & 0.7714     & 0.9145     & 84.166      \\
2  & outdex                              & 0.7476     & 0.9089     & 82.000     & MI                                  & ---     & ---     & ---     \\
3 & sakuraSAT                           & 0.7460     & 0.8966     & 80.333     & sakuraSAT                           & 0.7603     & 0.9123     & 82.7381     \\
4  & nakedGun                            & 0.7458     & 0.8990     & 81.066     & Ninofiero                           & 0.7525     & 0.9065     & 81.190      \\
5  & Ninofiero                           & 0.7442     & 0.8965     & 80.333     & nakedGun                            & 0.7505     & 0.9141     & 82.618      \\
6  & MI                                  & 0.7430     & 0.8992     & 81.066     & outdex                              & 0.7476     & 0.9011     & 82.261     \\ \hline
---  & Baseline                                  & 0.739     & 0.9056     & 80.83     & ---                              & ---     & ---     & ---     \\ \hline
\end{tabular}\vspace{-0.3cm}
\end{table}

Results from Table \ref{tab:result} also reveal interesting observations. The performance of \emph{unocanda} and \emph{sakuraSAT} is stable because they obtain the same rank on the both sets. \emph{outdex} is the second best on the development set but obtains a poor result in the test set. This is because the method of \emph{outdex} may be overfitting on the training set. Conversely, \emph{MI} achieves the second rank in the test set while this team is in $6^{th}$ on the development set. The method of other teams, i.e. \emph{nakedGun} and \emph{Ninofiero} gets stable performance.\vspace{-0.2cm}

\section{Conclusion}\label{sec:conclusion}\vspace{-0.2cm}
This paper presents our efforts to address a re-ranking problem in Discovery Challenge: Learning to Re-Ranking Questions for Community Question Answering competition. In order to do this, SVMRank is selected by analyzing a set of L2R methods. To improve the performance, the selected method is optimized by using a search algorithm to find out appropriate hyperparameters. The method with the best hyper-parameters is applied in the development and test set. By using SVMRank, our method obtains the third on the development and the second on the test set of the competition. Experimental results conclude that SVMRank is efficient in ranking questions for community question answering.

For feature direction, a feature selection should be considered to improve the performance of SVMRank. Also, combining SVMRank with other L2R methods in the form of ensemble learning should be considered. Finally, using a user-defined kernel, e.g. tree kernel should be integrated into the current model.

%
%

\bibliographystyle{plain}
\bibliography{reference}

\end{document}